\documentclass[preprint]{aastex701}

\newcommand{\sdo}{{\it SDO}} %
\newcommand{\hmi}{{\it SDO}/HMI} %
\newcommand{\ms}{m\,s$^{-1}$} %
\usepackage{gensymb}
\usepackage{hyperref}
\usepackage{float}
\usepackage{placeins}  
\usepackage{soul}

\begin{document}
\title{Long-Term Clustering Pattern of Solar Active Regions and Their Potential Connection with Magneto-Rossby Waves}


\author[0000-0002-6308-872X]{Junwei Zhao}
\affiliation{W.~W.~Hansen Experimental Physics Laboratory, Stanford University, Stanford, CA 94305-4085, USA}
\email{junwei@sun.stanford.edu}

\author[0000-0002-2632-130X]{Ruizhu Chen}
\affiliation{W.~W.~Hansen Experimental Physics Laboratory, Stanford University, Stanford, CA 94305-4085, USA}
\email{rzchen@stanford.edu}

\author[0000-0003-2622-7310]{Aimee A. Norton}
\affiliation{W.~W.~Hansen Experimental Physics Laboratory, Stanford University, Stanford, CA 94305-4085, USA}
\email{aanorton@stanford.edu}

\begin{abstract}
Large solar active regions (ARs) tend to be long lived and spatially clustered, with repeated emergence occurring in persistent solar activity nests over extended timescales.
By analyzing long-term spatiotemporal magnetic flux maps constructed from near-side magnetic field observations and far-side helioseismic AR maps, we investigate the recurrence and clustering properties of large ARs during Solar Cycle 24.
We find that, in both hemispheres, over 63\% of magnetic fluxes emerge and cluster within or near three distinct bands in the spatiotemporal maps, two of which rotate faster than the Carrington rate and one more slowly. 
These bands closely correspond to low-order nonaxisymmetric modes, primarily the azimuthal order $m$=1 mode. 
The drift rates of the three spatiotemporal bands are in good agreement with the phase speeds inferred for these modes.
The frequencies of the dominant modes are consistent with slow magneto-Rossby waves originating in the solar tachocline, associated with odd harmonic degrees $\ell$ and a toroidal magnetic field strength of approximately 4.0\,kG.
Our results suggest that magneto-Rossby waves play an important role in modulating both the timing and longitudinal localization of major AR emergence. 
Rieger-type periodicities may arise from interactions between a dominant mode and weaker modes, while longer quasi-periodic variations on 0.6--4\,yr timescales are likely linked to intersections of multiple major modes.
These findings point to a potential connection between surface magnetic flux patterns and dynamical processes in the tachocline.
\end{abstract}


\section{Introduction}
\label{sec1}

It has long been recognized that solar active regions (ARs) tend to recur at similar longitudes over consecutive rotations, often persisting for many months or even years. 
These longitudinal concentrations, variously referred to as activity nests, active longitudes, or hot spots, represent one of the most persistent nonaxisymmetric signatures of solar magnetic activity. 
\citet{Carrington1863} was the first to note the recurrence of preferred longitudes, and subsequent analyses of extended datasets across multiple observables further strengthened this picture \citep[e.g.,][]{Bumba1969, Bogart1982, Castenmiller1986, Bai1987, Benevolenskaya1999, detoma2000, Bai2003}. 
Later studies established that these active longitudes do not remain fixed in the Carrington frame but instead drift or rotate at slightly different rates. 
For example, \citet{Ruzmaikin2001} analyzed the lowest-order nonaxisymmetric magnetic modes and identified a rotation period of 27.03 days, faster than both the equatorial surface rate and the internal rotation profile, yet consistent with the rotation of large-scale solar-wind structures. 
\citet{Berdyugina2003} further showed that active longitudes often appear in antipodal pairs separated by roughly $180\degr$, persist for more than a century, and migrate at variable rates over the solar cycle. Alongside significant nest flux at rotation rates associated with differential rotation,
\citet{Norton2025} found that significant amount of activity nest fluxes rotated with a synodic 451--452 nHz prograde rate and a 409--411 nHz retrograde rate. 

In addition to the tendency of longitudinal clustering, AR emergences exhibit quasi-periodic modulation on timescales from several months to a few years. 
The $\sim$150--160-day Rieger-type periodicity was first identified in the recurrence of major solar flares \citep{Rieger1984}, and was later detected in flare rates, CME occurrence, and sunspot areas, all showing similar periodic behavior \citep[e.g.,][]{Bogart1985, Bai1987, Lean1990, Bai2003, Lou2003, Xiang2021}. 
Longer-period variations, commonly referred to as quasi-biennial oscillations (QBOs), span roughly $0.6 - 4$\,yr and appear in numerous magnetic and radiative activity indices. 
Examples include the 1.3-yr periodicity detected near the tachocline in internal rotation measurements \citep{Howe2000}, the $\sim$2-yr modulation in $p$-mode frequencies during Solar Cycle 24 \citep{Jain2023}, and the identification of QBOs as fundamental timescales of global magnetic-field variability \citep{Vecchio2012}. 
A more complete overview of additional findings about QBOs can be found in the review by \citet{Bazilevskaya2014}.

Efforts to connect the Rieger-type periodicities and quasi-biennial oscillations (QBOs) with Rossby waves date back several decades, motivated by the close agreement between the observed timescales and those expected for equatorially trapped Rossby modes. 
\citet{Gilman1969} first established the theoretical framework for solar Rossby waves, and \citet{Lou2000} examined the periods of equatorial Rossby and mixed Rossby-Poincar\'{e} waves, showing that several modes align well with the Rieger-type periodicities. 
Subsequent studies explored whether magneto-Rossby waves in the thin tachocline layer could account for the observed quasi-periodicities. 
For instance, \citet{Zaqarashvili2010a} demonstrated that cyclic variations of the toroidal magnetic field can selectively amplify harmonics whose periods fall within the Rieger range, while \citet{Zaqarashvili2010b} showed that the combined effect of differential rotation and a strong toroidal field ($> 10^5$\,G) can destabilize the $m=1$ harmonic with a characteristic period of $\sim$2\,yr.

These ideas gained renewed momentum following the direct detection of global-scale Rossby waves, through tracking coronal bright points \citep{McIntosh2017} and through helioseismic measurements of flow vorticity \citep{Loptien2018, Liang2019, Waidele2023}. 
Using an MHD shallow-water tachocline model, \citet{Dikpati2018} computed the phase speeds of magnetized Rossby waves and found good agreement with observed propagation speeds in the corona and coronal holes \citep{McIntosh2017, Krista2018}. 
\citet{Dikapti2018b} also investigated the role of inertial waves in generating solar seasons through studying the interaction between Rossby waves and differential rotation, and later they also studied how Rossby waves helped to decipher the deep origins of active regions \citep{Dikpati2021}. 
\citet{Korsos2023} applied wavelet analysis to synoptic magnetic maps and identified oscillatory signatures with Rossby-like periods, and \citet{Raphaldini2023} examined the longitudinal drift of magnetic patterns, finding drift speeds consistent with magnetic Rossby waves in a tachocline containing a 5--10\,kG toroidal field. 
Together, these results increasingly suggest that photospheric magnetic quasi-periodicities, from a few months to several years, may originate from magneto-Rossby or Rossby waves operating in the deep solar interior, likely near the tachocline.

Despite progress, most prior work focused on matching observed periodicities to theoretical mode periods for selected magnetic-field strengths, leaving unresolved questions about where and when ARs emerge and how they cluster to form active longitudes. 
Using a unique dataset that includes daily near-side magnetograms together with daily far-side magnetic-flux reconstructions, we are able to characterize the recurrence and organization patterns of solar magnetic fields over extended timescales. 
Fourier analysis of these emergence patterns reveals the dominant prograde and retrograde frequencies associated with low-degree magneto-Rossby waves, and these frequencies match well with a magneto-Rossby wave model with a toroidal field strength around 4\,kG in the tachocline.
This paper is organized as follows. 
Section~\ref{sec2} describes the data sources and processing. 
Section~\ref{sec3} presents the magnetic-flux integration analysis and the resulting organization patterns of major activity nests. 
Section~\ref{sec4} reports the Fourier analysis and the dominant prograde and retrograde frequencies for low-$m$ modes. 
We discuss the implications of these results and their connection to the interior magneto-Rossby waves in Section~\ref{sec5}, and give conclusions in Section~\ref{sec6}.

\section{Data Preparation}
\label{sec2}

\begin{figure}[!t]
    \centering
    \includegraphics[width=0.55\linewidth]{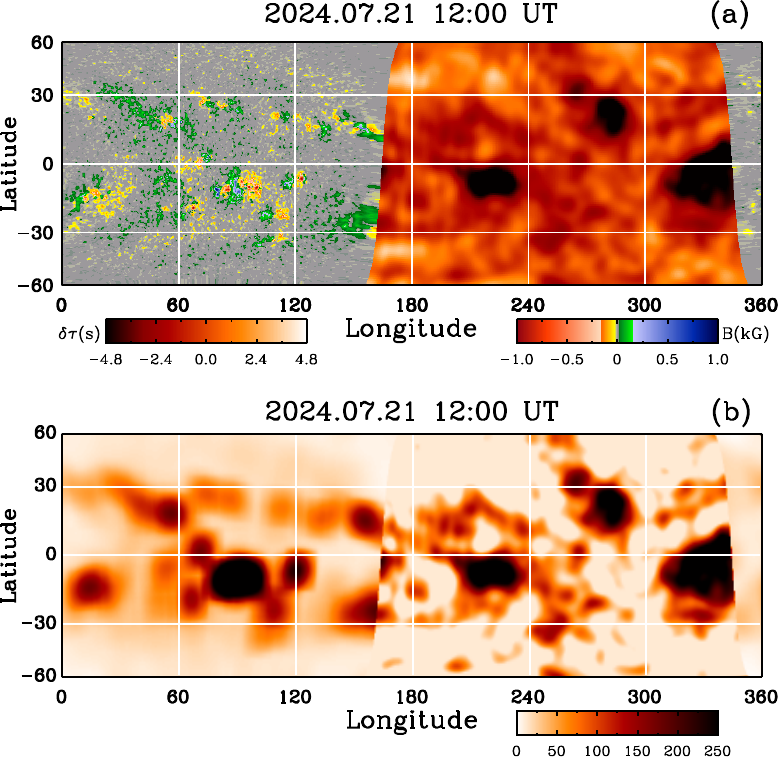}
    \caption{(a) Synchronic image of the Sun for 12:00 UT of 2024 July 21, combining near-side magnetic-field observations (green–orange on a gray background) and far-side helioseismic AR images (dark patches on an orange background). 
    The color bars indicate the ranges of magnetic-field strength and acoustic travel-time deficits, separately.
    (b) Synchronic image of ARs, derived from panel (a) after converting the near-side magnetic field to magnetic flux and degrading the spatial resolution to match that of the helioseismic far-side images.
    The map is displayed after all values are rescaled to $0-255$, with the maximum value corresponding to, approximately, magnetic flux strength of 50\,G after the resolution degradation.}
    \label{fig:synchronic}
\end{figure}

The data used in this analysis include both near-side magnetic-field observations from Helioseismic Magnetic Imager \citep[HMI;][]{Scherrer2012, Schou2012} onboard {\it Solar Dynamics Observatory} \citep[\sdo;][]{Pesnell2012} and far-side helioseismic images of active regions (ARs) that are derived from \hmi\ Dopplergrams using time--distance helioseismology \citep{Zhao2019, Chen2022}. 
At any given time, only half of the Sun is directly visible to the ground-based or near-Earth observing instruments, and helioseismic far-side imaging technique offers complementary images to construct synchronic full-Sun representations of solar ARs. 
Figure~\ref{fig:synchronic}a shows an example of such a synchronic image, taken on 2024 July 21 when the Sun was magnetically active on both sides.

However, as clearly shown in Figure~\ref{fig:synchronic}a, the near-side magnetic-field observations have high spatial resolution and capture both magnetic polarities, whereas the far-side helioseismic images show only acoustic travel-time deficits, with substantially lower spatial resolution and without information of magnetic polarities. 
To construct a multi-year dataset with consistent AR properties across the full sphere, we convert the near-side magnetic-field maps into magnetic-flux maps by dropping the polarity signs, and then convolve the resulting maps with a two-dimensional Gaussian function to degrade their spatial resolution and match that of the helioseismic far-side ARs.
The spatial resolution for far-side ARs is generally believed to be around $10\degr$, corresponding to the characteristic scales of the dominant modes used in these helioseismic far-side imaging technique.  
To match the resolution of far-side images, we choose the two-dimensional Gaussian function, which is used for the near-side image degradation, to have an FWHM of $15\fdg0$ in both longitudinal and latitudinal directions. 

Figure~\ref{fig:synchronic}b shows the resulting synchronic magnetic-flux map obtained from Figure~\ref{fig:synchronic}a after dropping the polarities and blurring the near-side observations. 
Although the near-side portion still apparently retains better image quality than the far-side portion, the overall AR properties appear more comparable on both sides. 
It can also be seen that after the Gaussian smoothing, the ARs cover substantially larger areas than where magnetic fields are, but this is not expected to impact our results because in this work we will analyze large-scale magnetic structures. 
Both the near-side and far-side images are rescaled to the range of 0-255 so that they can be plotted together, and the maximum value corresponds to, approximately, magnetic flux strength of 50\,G after the Gaussian smoothing is applied.
Such low-resolution images are used as magnetic-flux proxies in the analysis that follows; however, for simplicity but without introducing any potential misunderstanding, we use the term ``magnetic flux" to describe the quantities shown in such images.

\begin{figure}[!t]
    \centering
    \includegraphics[width=0.75\linewidth]{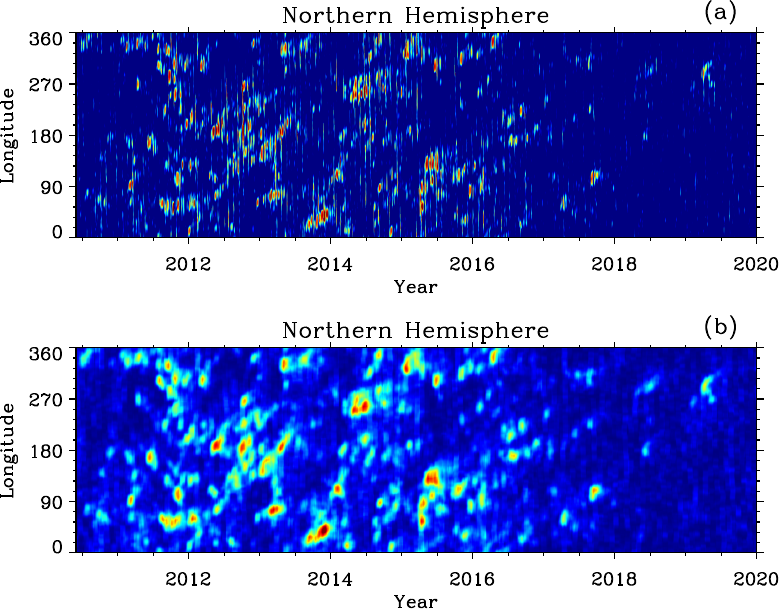}
    \caption{(a) Spatiotemporal map of magnetic-flux proxies for the northern hemisphere, constructed by taking daily synchronic images, collapsing the flux within the $0\degr$--$30\degr$N latitudinal band, and stacking the resulting curves over the 3500-day course. 
    The magnetic-flux values are in arbitrary units due to the combination with far-side helioseismic data and the smoothing applied.
    (b) Spatiotemporal image for the northern hemisphere after a 27-day running smoothing is applied on the data shown in panel (a). }
    \label{fig:long_sequence_N}
\end{figure}

In this study, we investigate the long-term emergence and clustering patterns of ARs throughout the \hmi-observed period of Solar Cycle 24, covering 2010 June 1 to 2019 December 30.
Note that the full period of Cycle 24 is approximately from December 2008 to December 2019, and this set of observational data missed the first 18 months of the cycle.
But due to the low activity level at the beginning of the cycle, this study is not much impacted. 
Also, the \hmi\ observations covered Solar Cycle 25 from 2020 January to the present, but because the cycle will likely continue for a few more years, the days in Cycle 25 are thus not included in the current analysis.

\begin{figure}[!t]
    \centering
    \includegraphics[width=0.75\linewidth]{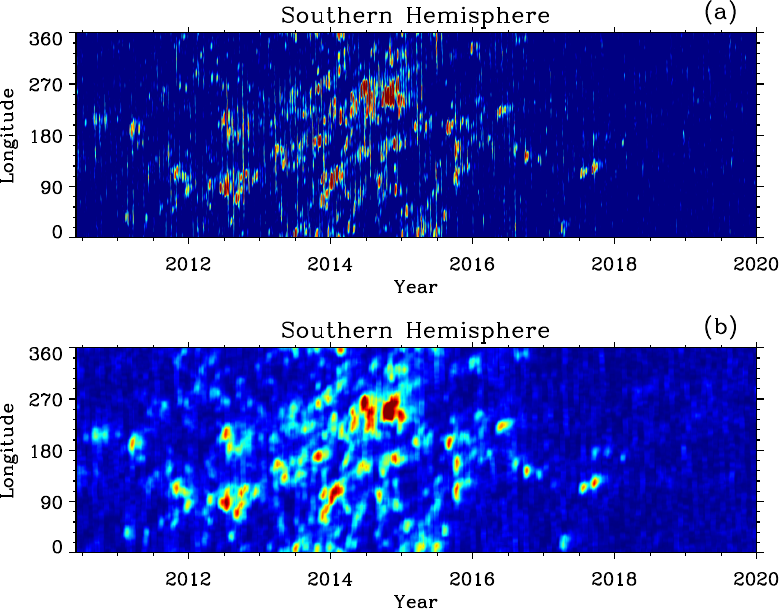}
    \caption{Same as Figure~\ref{fig:long_sequence_N} but for the southern hemisphere. }
    \label{fig:long_sequence_S}
\end{figure}

To prepare the daily synchronic images for this study, one synchronic magnetic-flux image, as the example shown in Figure~\ref{fig:synchronic}b, is taken for each calendar day and a total of 3500 images spanning over 9.5 years are collected. 
In the following analysis, for each synchronic image we collapse the magnetic flux within the $0\degr$--$30\degr$N latitudinal band into a longitude-dependent curve for the northern hemisphere, and flux within $0\degr$--$30\degr$S is collapsed to form a separate curve for the southern hemisphere. 
The collections of these daily curves form a two-dimensional (time and Carrington longitude) spatiotemporal map for each hemisphere (see Figure~\ref{fig:long_sequence_N}a for the northern hemisphere and Figure~\ref{fig:long_sequence_S}a for the southern hemisphere).

Similar spatiotemporal maps based on synoptic magnetic charts have been analyzed by various authors \citep[e.g.,][]{Raphaldini2023}, but the inclusion of helioseismic far-side images allows us to construct the long-term spatiotemporal maps with a temporal cadence of 1 day rather than 1 Carrington rotation (approximately, 27.3 days), greatly enhancing the temporal cadence for our analysis in the time domain. 
As seen in Figures~\ref{fig:long_sequence_N}a and \ref{fig:long_sequence_S}a, some vertical stripes appear in the maps due to occasional strong fluctuations on the far-side images. 
To suppress these artifacts and improve the overall smoothness, we apply a 27-day running smoothing in the temporal direction of the maps, and the chosen window length corresponds approximately to one Carrington rotation. 
Note that spatial smoothing along  longitudes has already been applied in an earlier step through Gaussian convolution. 
Figures~\ref{fig:long_sequence_N}b and \ref{fig:long_sequence_S}b show the resulting smoothed spatiotemporal maps for the northern and southern hemispheres, respectively.

A total of approximately 2500 NOAA-designated ARs appeared on the Sun during Solar Cycle 24, and nearly all of them (excluding those appearing above $30\degr$ latitude) are included in the averaged and smoothed flux maps of Figures~\ref{fig:long_sequence_N} and ~\ref{fig:long_sequence_S}. 
These maps are therefore a valuable asset for examining the AR emergence and clustering patterns over the course of the cycle.

\section{Emergence and Clustering Patterns of Active Regions}
\label{sec3}

It is evident from Figures~\ref{fig:long_sequence_N} and \ref{fig:long_sequence_S} that large and long-lived  ARs do not appear randomly on the solar surface, but tend to cluster in space and time. 
However, they do not cluster along one or more Carrington longitudes; instead, these ARs tend to drift to neighboring longitudes and appear persistently for a few months.
Recognizing that Carrington rotation rate is simply one arbitrarily chosen rate within the Sun’s broad range of differential rotation rates, we are motivated to examine whether ARs tend to align more preferably 
when they are viewed under different rotation rates.

\begin{figure}
    \centering
    \includegraphics[width=0.98\linewidth]{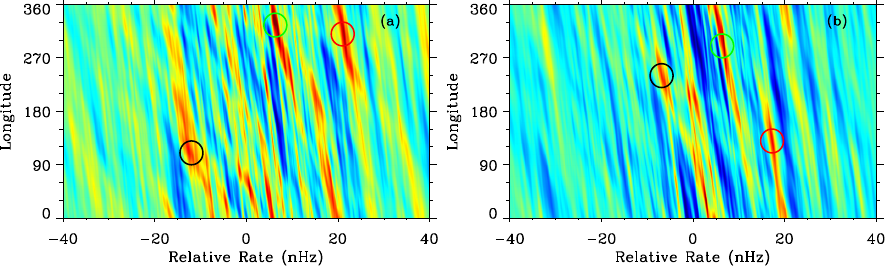}
    \caption{Total flux integrated within $20\degr$-wide bands that have different tracking rates relative to the Carrington rotation rate and have different initial longitudes. 
    Display scales are arbitrary in both panels, with deep blue corresponding to minimum total flux and red indicating maximum total flux. 
    Positive values of the relative rate indicate prograde rate and negative values retrograde.
    Vertical axis indicates the longitude where the middle of the integration band starts at the beginning time.
    Panel (a) shows results for the northern hemisphere calculated from Figure~\ref{fig:long_sequence_N}b, and panel (b) displays results for the southern hemisphere calculated from Figure~\ref{fig:long_sequence_S}b.
    In each panel, three circles in colors of black, green, and red (from left to right) show the locations with three strongest flux concentrations, and these three locations are overplotted in the same colors as slanted bands in Figures~\ref{fig:connct_lines_N} and \ref{fig:connct_lines_S}.
    }
    \label{fig:hough}
\end{figure}

Figures~\ref{fig:long_sequence_N}b and \ref{fig:long_sequence_S}b show the daily magnetic-flux distribution as a function of Carrington longitude, meaning that the data are plotted with a tracking rate equal to the Carrington rotation rate. 
Any rotation rates other than the Carrington rate correspond to slanted lines relative to the horizontal axis on these spatiotemporal maps. 
For each tracking rate from $-40$ to $40$\,nHz relative to the Carrington rate, we integrate for the total magnetic flux along a slanted band of $20\degr$ wide ($\pm10\degr$ from the middle line of the band) over the full time period. 
For each integration band, another parameter in addition to the tracking rate is the offset on the vertical axis (longitude), which varies from $0\degr$ to $360\degr$. 
The middle line of the band is selected to correspond to this vertical offset. 
An integrated flux value can thus be obtained for each tracking rate and each initial longitude, and these flux values form a two-dimensional map for each hemisphere (Figure~\ref{fig:hough}).

\begin{figure}
    \centering
   \includegraphics[width=0.80\linewidth]{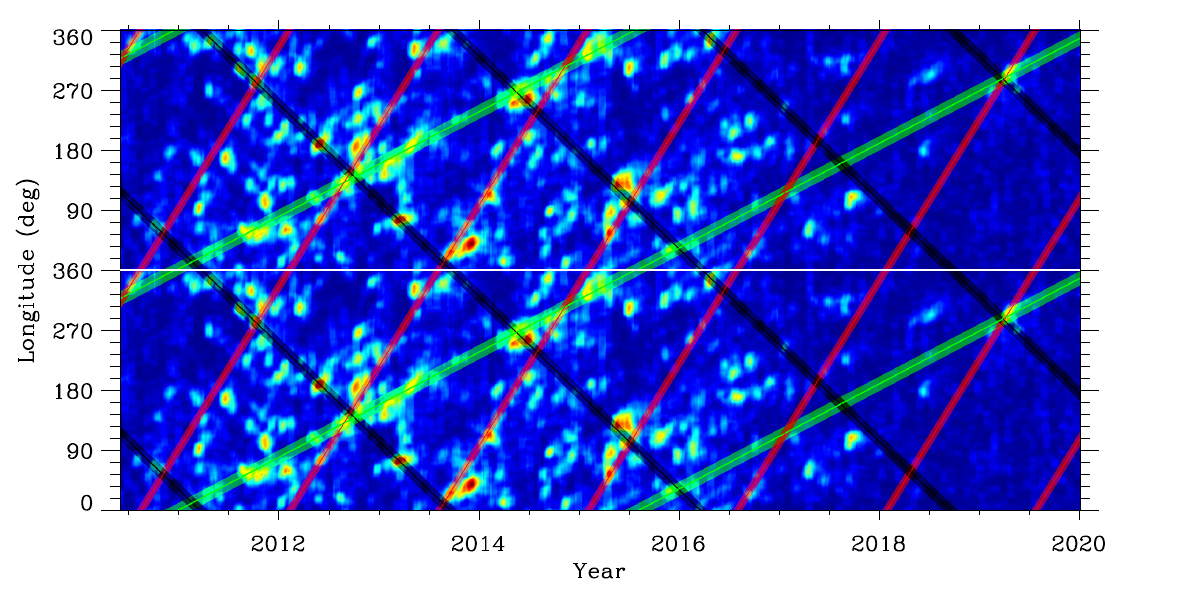}
    \caption{Three integration bands are overplotted on the spatiotemporal maps of the northern hemisphere. 
    These bands correspond to the three strongest flux-concentration regions identified in Figure~\ref{fig:hough}a, plotted in the same colors as their corresponding circles in that figure. 
    The green, red, and black bands represent prograde drift rates of $5\fdg9 \pm 0\fdg3$ and $17\fdg2 \pm 0\fdg3$ per Carrington rotation, and a retrograde rate of $10\fdg8 \pm 0\fdg3$ per rotation, respectively.
    To illustrate the continuity of these lines, two copies of the spatiotemporal map (same as in Figure~\ref{fig:long_sequence_N}b) are stacked vertically. 
     For clarity, some bands are omitted in the later part of the cycle when few ARs appear.
   }
    \label{fig:connct_lines_N}
\end{figure}

\begin{figure}
    \centering
    \includegraphics[width=0.80\linewidth]{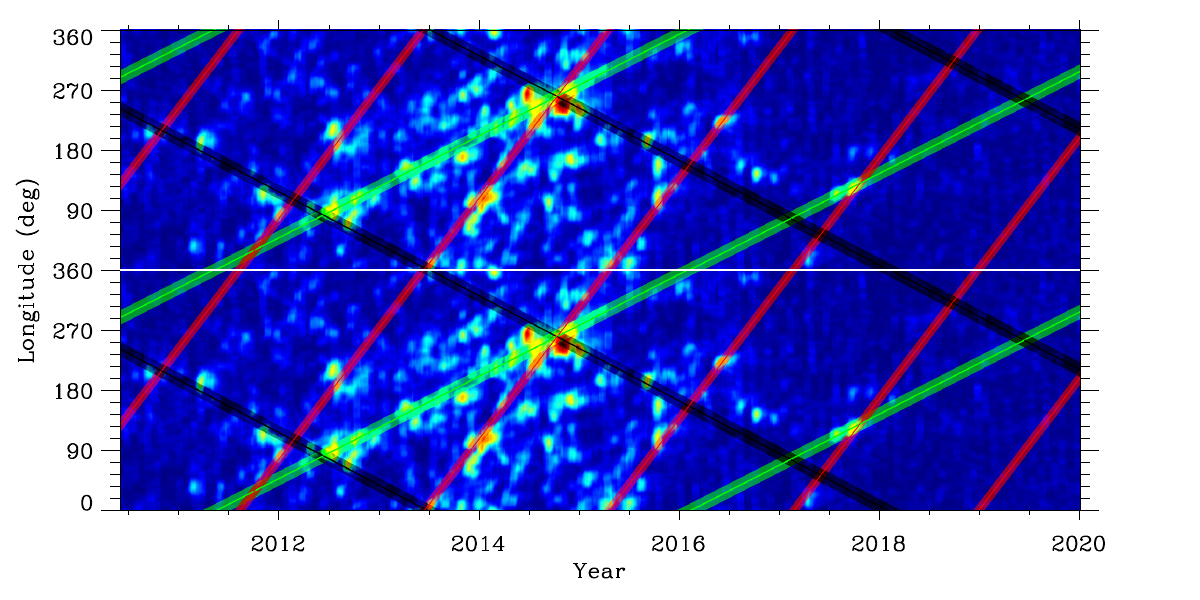}
    \caption{Same as Figure~\ref{fig:connct_lines_N} but for the southern hemisphere.
    The green, red, and black bands represent prograde drift rates of $5\fdg7 \pm 0\fdg3$ and $14\fdg6 \pm 0\fdg2$ per Carrington rotation, and a retrograde rate of $5\fdg9 \pm 0\fdg3$ per rotation, respectively.}
    \label{fig:connct_lines_S}
\end{figure}

As seen in either panel of Figure~\ref{fig:hough}, several areas of enhanced integrated flux appear, indicating that a few specific tracking rates yield significantly larger total flux than other rates or Carrington rate, which is 0 in the plot. 
This suggests that there may be a few internal rates, prograde or retrograde, that better characterize when and where large amounts of magnetic fluxes emerge. 
For each hemisphere, we identify three (note that the choice of three is arbitrary, but a larger number will make the follow-up analysis much messier) strongest flux concentrations, encircled in three different colors in Figure~\ref{fig:hough}, and overplot their corresponding $20\degr$-wide integration bands in the same color on the spatiotemporal maps to better visualize the AR-clustering patterns (Figures~\ref{fig:connct_lines_N} and \ref{fig:connct_lines_S}).
Note that the flux concentrations in Figure~\ref{fig:hough} often cover large areas, and we choose the relative rate corresponding to the highest concentrations for plotting the integration bands. 

Figure~\ref{fig:connct_lines_N} presents the northern-hemisphere results. 
In order to better visualize the continuation of the slanted bands, we stack the same spatiotemporal map twice vertically in the figure, and each band, depending upon their slopes (i.e., relative speeds), repeat different number of times across the spatiotemporal maps. 
It can be seen that not all ARs fall into the overplotted bands, but most of the large and strong ARs (with reddish color and larger sizes) are either within the bands or are close to the bands.
Our estimation shows that around 63\% of the total magnetic fluxes locate either within the three bands or within $20\degr$ distance from these bands.
It is also notable that the few largest ARs of the whole cycle are located in the vicinity of the intersections of two or three bands.

Among the three slanted bands that yield the highest integration fluxes, two bands show prograde rates relative to Carrington rate and one shows retrograde rate. 
The green, red, and black lines correspond to relative tracking rates and initial longitudes of ($6.9 \pm 0.4 $\,nHz, $315\degr$), ($20.3 \pm 0.3$\,nHz, $310\degr$), and ($-12.7 \pm 0.4$\,nHz, $115\degr$), respectively. 
The uncertainties in the rates are estimated through the sizes of the flux concentrations in Figure~\ref{fig:hough}.
These tracking rates translate into prograde drift rates of $5\fdg9 \pm 0\fdg3$ and $17\fdg2 \pm 0\fdg3$ per Carrington rotation, and a retrograde rate of $10\fdg8 \pm 0\fdg3$ per rotation, which again correspond to drift speeds of $29.1 \pm 1.7$\,\ms, $85.9 \pm 1.3$\,\ms, and $-53.7 \pm 1.7$\,\ms, respectively.

Figure~\ref{fig:connct_lines_S} shows the results of the southern hemisphere. 
Again, three slanted bands represent the tracking rates producing the highest integrated fluxes, and they correspond to relative rotational rates and initial longitudes of ($6.7 \pm 0.3$\,nHz, $287\degr$), ($17.2 \pm 0.2$\,nHz, $127\degr$), and ($-6.9 \pm 0.4$\,nHz, $240\degr$). 
These relative rates translate into prograde rates of $5\fdg7 \pm 0\fdg3$ and $14\fdg6 \pm 0\fdg2$ per Carrington rotation, and a retrograde rate of $5\fdg9 \pm 0\fdg3$ per rotation. 
These relative rates are equivalent to drift speeds of $28.3 \pm 1.3$\,\ms, $72.8 \pm 0.8$\,\ms, and $-29.2 \pm 1.7$\,\ms, respectively.
Note that the first drift speeds for both the northern and southern hemispheres, corresponding to the green bands in the spatiotemporal maps, are remarkably similar.

From Figures~\ref{fig:connct_lines_N} and \ref{fig:connct_lines_S}, it is clear that in both hemispheres, the majority of large, long-lived ARs lie within or near three dominant bands, and the strongest AR clusters of the entire solar cycle tend to occur within or near the intersections of two or three of these bands.
These indicate that major ARs tend to appear and cluster in a pattern that seems to be determined by a few straight bands.

\section{Power Spectrum of Magnetic Fluxes}
\label{sec4}

The integration of total fluxes using different rotation rates in the previous section shows that magnetic flux tends to accumulate preferentially with certain rotation rates. 
This motivates us to examine whether the emergence and clustering of major ARs are related to magneto-Rossby waves in the deep solar interior,as previously proposed by various authors through analyzing organizing patterns of ARs \citep[e.g.,][]{Dikpati2018, Korsos2023, Raphaldini2023} or coronal bright point densities \citep{Raphaldini2024}. 
In particular, it is worth exploring whether magneto-Rossby waves, occurring in the solar interior or tachocline with preferred frequencies (or periods) and azimuthal orders $m$ (or spatial scales), could modulate or influence the timing and location of the emergence and recurrence of major ARs, thereby imprinting signatures of those waves.

\begin{figure}
    \centering
    \includegraphics[width=0.80\linewidth]{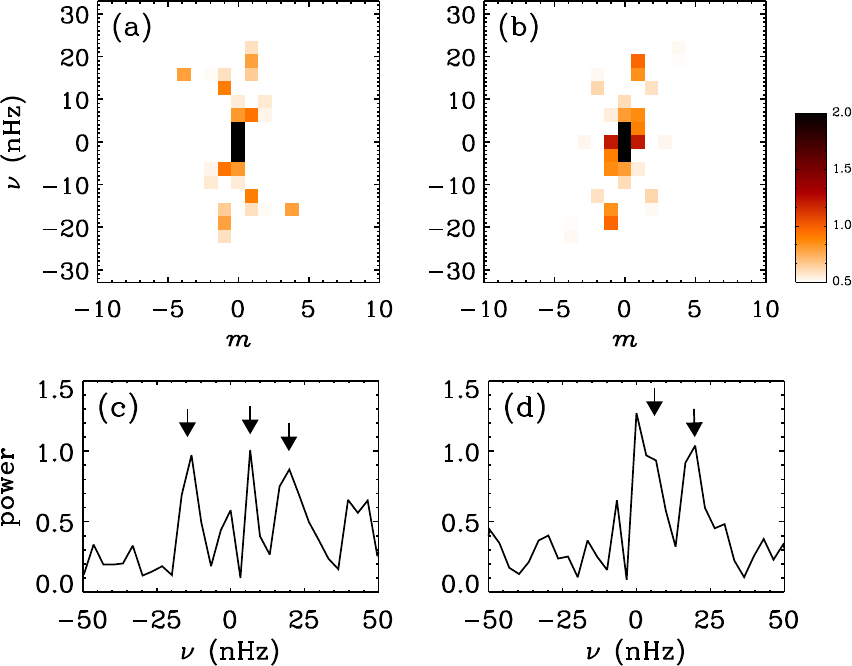}
    \caption{(a) Power spectrum calculated from the spatiotemporal magnetic-flux map (Figure~\ref{fig:long_sequence_N}b), obtained from the latitudinal band of $0\degr-30\degr$N.
    (b) Same as panel (a) but for the southern hemisphere. 
    (c) Power spectrum of the northern hemisphere for $m$=1, displayed as a function of $\nu$. 
    Arrows point to the dominant modes.
    (d) Same as panel (c) but for the southern hemisphere. 
    Note that the unit for power is arbitrary. }
    \label{fig:pow_NS}
\end{figure}

To investigate this possibility, we compute the power spectra of the spatiotemporal magnetic-flux maps for the northern and southern hemispheres separately. 
We perform a two-dimensional Fourier transform on the data shown in Figures~\ref{fig:long_sequence_N}b and \ref{fig:long_sequence_S}b, obtaining the power as a function of azimuthal order $m$ and frequency $\nu$, which are the Fourier components corresponding to longitude and time, respectively. 
In the Fourier domain, the waves follow a function $\propto \exp[i 2\pi(m\phi - \nu t)]$, where $\phi$ represents longitude and $t$ time.
Under this definition, positive $m$ and positive $\nu$ represent prograde modes, and positive $m$ and negative $\nu$ represents retrograde modes.
The negative-$m$ side of the power spectra provides the same information as the positive-$m$ side.

Figure~\ref{fig:pow_NS}a shows the power spectrum for the $0\degr-30\degr$N band. 
As seen in this figure panel, two prograde modes are prominent with some frequency spread at ($m$=1, $\nu= 6.6 \pm 1.6$\,nHz) and ($m$=1, $\nu= 19.7 \pm 4.2$\,nHz), and one prominent retrograde mode appears at ($m$=1, $\nu= -14.6 \pm 3.2$\,nHz). 
There is another prominent mode, ($m$=4, $\nu= -16.5 \pm 1.6$\,nHz), but in this study, we limit our discussions to only the three strongest signals.
The frequencies and their spreading ranges are estimated from the power-weighted averages and power-weighted deviation of all signals near the strongest signals, and we do not consider contributions from $\nu=0$ when making these calculations.
These three dominant modes correspond to phase velocities ($v_\mathrm{ph} = 2 \pi \nu R_\odot \cos \lambda / m$, where $\lambda$ is latitude and we take $15\degr$ here) of, relative to the co-rotating Carrington rotation frame, $27.9 \pm 6.8$\,\ms, $83.2\pm17.7$\,\ms, and $-61.7\pm 13.5$\,\ms, respectively. 
Unsurprisingly, these modes correspond to the three dominant bands plotted in Figure~\ref{fig:connct_lines_N}, with their phase speeds in reasonable agreement with the drift speeds of those bands estimated earlier in Section~\ref{sec3}.



Figure~\ref{fig:pow_NS}b presents the results for the southern hemisphere. 
Again, two clusters of prograde power appear at $m=1$, with power-weighted frequencies at $\nu= 6.1 \pm 1.7$\,nHz and $19.4 \pm 2.3$\,nHz. 
One retrograde mode appears at ($m$=2, $\nu=-14.7 \pm 2.5$\,nHz). 
These three modes have phase velocities of $25.7\pm7.2$\,\ms, $82.0\pm9.7$\,\ms, and $-31.1\pm5.3$\,\ms, respectively, relative to the Carrington rotation.
These three modes correspond to the dominant bands shown in Figure~\ref{fig:connct_lines_S}, with their phase speeds well matching the drift speeds of the bands. 

The one-to-one correspondence of the three dominant flux bands and the three dominant modes in each hemisphere implies that the emergence and clustering of the major ARs seem to have certain frequencies and spatial scales, both of which are often linked to waves, presumably, magneto-Rossby waves. 

\section{Discussion and Conclusion}
\label{sec5}

By analyzing the long-term spatiotemporal magnetic-flux maps, constructed from daily near-side magnetograms and far-side magnetic-flux proxies during the \hmi-observed period of Solar Cycle 24, we have investigated the emergence and clustering patterns of major ARs in longitude and time, or when and where major ARs appear on the solar surface.
Through integrating magnetic flux along bands with different tracking speeds, we identified three dominant bands in each hemisphere that maximize the total fluxes: two with prograde speeds and one with retrograde speed. 
Major ARs are found to lie within or near these bands, with the particularly active ARs located in the vicinity of the intersections of two or more bands. 
Fourier analyses of the same spatiotemporal maps also reveal two dominant prograde modes and one dominant retrograde mode in each hemisphere, with the phase speeds of the modes closely matching the drift speeds inferred from the flux-integration method.

Before discussing the physical implications of our results, it is useful to note several methodological considerations. 
The use of low-resolution near-side observations together with far-side helioseismic images inevitably enlarges the apparent areas of magnetic regions relative to their true sizes. 
However, because our analysis focuses only on the large-scale patterns, corresponding to low-$\ell$ and low-$m$ spatial components, the reduced spatial resolution is not expected to influence the final results.
The omission of magnetic polarity information is necessary, as our analysis emphasizes the clustering patterns of magnetic flux rather than the detailed spatial distribution of magnetic fields. 
Analyses similar to those presented in this study could also be performed using synoptic magnetic-field maps, albeit with coarser temporal frequency resolution. 
Nevertheless, this limitation should not affect the primary conclusions, because the frequencies examined in this study lie well within the frequency range prescribed by the temporal cadences of synoptic maps. 
We plan to expand our analysis to include previous solar cycles using synoptic magnetic maps in a future study. 

\subsection{AR Clustering Patterns and Quasi-periodic Behaviors}
\label{sec51}

Previous studies have measured AR drift speeds by tracking long- or short-term AR motions \citep[e.g.,][]{Berdyugina2003, Raphaldini2023}. 
Those measured values largely correspond to the slower one of the two prograde drift speeds in our analysis, denoted by the green bands in Figures~\ref{fig:connct_lines_N} and \ref{fig:connct_lines_S}. 
As can be evidently seen in those figures, this drift pattern is the most prominent and persistent one throughout the solar cycle in both hemispheres. 
The nearly identical drift speeds in both hemispheres are unlikely to be coincidental; 
they are likely due to the constant faster rotational speeds of magnetic fields relative to the quiet Sun, consistent with the claim by \citet{Berdyugina2003}.
However, this prograde drift is not the only pattern associated with when and where AR appear. 
Each hemisphere also exhibits another faster prograde drift and a retrograde drift that determine AR appearance time and location. 
Additional weaker drift components may still be identified in Figures~\ref{fig:connct_lines_N} and \ref{fig:connct_lines_S}, and they may also contribute to the AR appearance patterns, but we focus only on the three dominant drift speeds in this study.

Figures~\ref{fig:connct_lines_N} and \ref{fig:connct_lines_S} reveal a picture in which the location and timing of AR emergence are modulated by three dominant traveling waves (possibly together with several weaker ones), propagating either prograde or retrograde. 
Magnetic fields, which are generated by solar dynamo, are likely influenced by these waves and emerge preferentially when local wave amplitudes exceed some threshold. 
This naturally explains why ARs cluster along the three dominant drifting bands, locations where wave amplitudes peak and their interference with other less prominent waves may help enhance the probability to exceed the threshold of emergence. 
These modulations provide a plausible physical mechanism for the Rieger-type periodicities near the neighborhood of 150\,days, although in the present study we do not attempt to identify those weaker waves.

When two or more dominant waves intersect, constructive interference may substantially increase the modulation of subsurface magnetic fields, triggering enhanced AR emergence at those times and locations. 
This offers a natural explanation for the observed quasi-biennial oscillations (QBOs). 
For example, in the northern hemisphere (Figure~\ref{fig:connct_lines_N}), the prograde green band intersects the retrograde black band every
\begin{equation}
 \frac{360\degr}{(5.9\degr + 10.8\degr)/\mathrm{rotation}} \approx 1.61\,\mathrm{yr},
\end{equation}
the red band intersects the black band every 0.96\,yr, and the green and red bands intersect every 2.38\,yr. 
In the southern hemisphere, the corresponding values are 2.14\,yr, 1.25\,yr, and 3.02\,yr, respectively. 
All those values are consistent with the QBO timescales of 0.6--4\,yr introduced in Section~\ref{sec1}. 
Differences in the peaking phases between hemispheres may further shorten the global recurrence interval of AR emergence, producing even shorter periodicities.
We stress that in our opinion, it is these intersections that determine the quasi-periodic behaviors of the solar ARs, rather than the periods of the waves or modes suggested by some previous authors \citep[e.g.,][]{Lou2000, Zaqarashvili2010b}. 

Finally, we note that the straight integration bands used in Section~\ref{sec3} with constant drift speeds, rather than with time-varying speeds, are just an approximation. 
The actual drift speeds of magnetic flux may vary within one solar cycle, as well as from cycle to cycle, due to variations of the internal magnetic-field strength (see Section~\ref{sec52}) and other physical conditions. 
It is therefore reasonable to speculate that these three same bands will unlikely extend into the following solar cycles without any alterations, because during the long solar minimum years, the drift speeds will undoubtedly differ from the maximum years.
At present, Solar Cycle 25 just passed its activity maximum phase, and we will analyze its AR emergence patterns until the cycle ends.
It would be interesting to see how these two cycles agree and differ from each other in terms of the drift speeds of the dominant waves. 

\subsection{Magneto-Rossby Waves and Potential Connections with AR Patterns}
\label{sec52}   

In the $m-\nu$ power spectra derived from the spatiotemporal magnetic-flux maps of both hemispheres, we identify three dominant power concentrations: two prograde and one retrograde (Figures~\ref{fig:pow_NS}). 
The phase speeds associated with these dominant modes agree well with the predominant drift speeds identified in Figures~\ref{fig:connct_lines_N} and \ref{fig:connct_lines_S}, confirming that the primary trends visible in the spatiotemporal images correspond directly to the dominant modes in the power spectra.


If the dominant modes are indeed associated with magneto-Rossby waves, then the mode frequencies and phase speeds are governed by the magnetic field strength in the tachocline, as suggested by earlier studies \citep[e.g.,][]{Zaqarashvili2010a, Zaqarashvili2018}.
The dispersion relations for fast and slow magneto-Rossby waves, corresponding to the retrograde and prograde modes, respectively, were derived by \citet{Zaqarashvili2007}:
\begin{equation}
\frac{\nu_{\ell m}}{m} = - \frac{\Omega}{2 \pi \ell (\ell + 1)} \bigg( 1 \pm \sqrt{ 1 + \frac{v_A^2}{\Omega^2 R^2} \ell (\ell + 1)[\ell (\ell + 1) - 2 ]} \ \bigg),
\label{eq2}
\end{equation}
where $\nu_{\ell m}$ is the mode frequency at harmonic degree $\ell$ and azimuthal order $m$, $\Omega$ is the rotation rate, $R$ is the radius, $v_A = B_0/\sqrt{4 \pi \rho}$ is the Alfv\'en speed, $B_0$ is the toroidal magnetic field strength, and $\rho$ is the mass density.
These parameters describe the physical conditions of the solar interior in which magneto-Rossby waves arise.
But it is worth mentioning that this is a simplified equation with an assumption of rigid-body rotation ignoring the latitudinal and radial differential rotations, as well as an assumption of the global magnetic field being proportional to cosine of the latitude.

The `$+$' and `$-$' signs in Equation~\ref{eq2} indicate that, for each $(\ell, m)$, there exists a pair of modes corresponding to the slow and fast branches, respectively.
It is therefore tempting to pair two observed modes with different $\nu$ values but the same $m$, and use them to estimate the toroidal magnetic field strength in the solar interior.
Assuming that the magneto-Rossby waves originate in the tachocline, we adopt typical values at the depth of $0.72\,R_\sun$ (near the top of the tachocline) and latitude of $15\degr$:
$\Omega / 2 \pi = 441.4$\,nHz, $R = 5.0 \times 10^{10}$\,cm, and $\rho = 0.21$\,g\,cm$^{-3}$.
Note that the vertical axes in Figure~\ref{fig:pow_NS}, as well as the $\nu$ values reported in Section~\ref{sec4}, are defined relative to the Carrington rotation rate, whose sidereal value is 456.03\,nHz, significantly higher than the tachocline rotation rate.
We attempted to pair the dominant modes identified in Section~\ref{sec4} and solve for $B_0$, but were unable to find a solution that simultaneously satisfies the dispersion relations for both the slow and fast modes.

\begin{figure}[!t]
    \centering
    \includegraphics[width=0.85\linewidth]{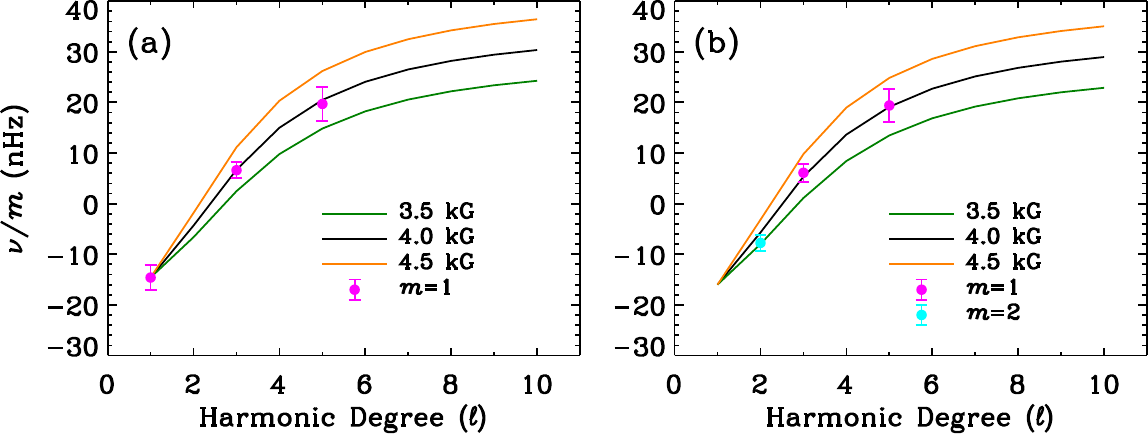}
    \caption{(a) Solid curves of three different colors show the normalized frequencies $\nu/m$ as functions of $\ell$ for different toroidal field strengths of 3.5, 4.0, and 4.5\,kG. 
    The calculation is made with typical physical conditions in the solar tachocline for the slow magneto-Rossby waves.
    Magenta points correspond to $\nu$ for the three dominant $m=1$ modes found in the northern hemisphere, as shown in Figure~\ref{fig:pow_NS}a. 
    (b) Same as panel (a) but for the southern hemisphere. 
    The cyan point corresponds to $\nu$ for the dominant $m=2$ mode.     }
    \label{fig:ell_power}
\end{figure}

However, we are able to find a satisfactory solution using a different approach.
Because all dominant modes reported in Section~\ref{sec4} are prograde relative to the tachocline, it is plausible that they all correspond to slow magneto-Rossby waves with different harmonic degrees $\ell$ but the same azimuthal order $m$=1 (except for one $m$=2 mode in the southern hemisphere).
Assuming $B_0 = 3.5$, 4.0, and 4.5\,kG in the tachocline, we compute $\nu/m$ for all $\ell < 10$.
As shown in Figure~\ref{fig:ell_power}a, the frequencies measured in the northern hemisphere for $m$=1 agree very well with the theoretical expectations for odd $\ell$'s ($\ell$=1, 3, 5) when $B_0 = 4.0$\,kG.
In the southern hemisphere, the two $m$=1 modes show similarly good agreement with the expected values for odd $\ell$'s ($\ell$=3 and 5).
The single $m$=2 mode identified in Section~\ref{sec4} also agrees with the theoretical prediction, but corresponds to an even $\ell$ ($\ell$=2).
In the southern hemisphere calculations, we adopt $\Omega / 2\pi = 440.0$\,nHz, which is slightly slower than the value used for the northern hemisphere.

The close agreement between the theoretical predictions and the measured frequencies provides strong evidence that the observed surface magnetic flux patterns are likely caused by modulations of slow magneto-Rossby modes originating in the tachocline.
Magneto-Rossby modes with different $(\ell, m)$ combinations can produce a wide range of frequencies, and their superposition likely gives rise to the complex yet quasi-periodic patterns of active region emergence and clustering.

We also emphasize that the measured mode frequency $\nu$ is highly sensitive to the toroidal magnetic field strength $B_0$ (see Figure~\ref{fig:ell_power}).
A change of only 0.5\,kG in $B_0$ can lead to a frequency shift of several nHz, indicating that precise measurements of mode frequencies can provide a robust constraint on the toroidal field strength near the tachocline.
The inferred value of $B_0 \approx 4.0$\,kG is comparable to the $5$--$10$\,kG estimate of \citet{Raphaldini2023} and is notably consistent with equipartition field strengths $B_\mathrm{eq}$ expected at the base of the convection zone. 
Mixing-length estimates for convective velocity yields $B_\mathrm{eq}$ values between $\sim$3\,kG \citep{Brandenburg2005} and $\sim$10\,kG \citep{Fan2021}, and our inferred field strength is well within that range. 
In addition, \citet{Brandenburg2005} noted that an order-of-magnitude estimate based on the total unsigned magnetic flux emerging at the solar surface over a cycle yields a mean field of around $4$\,kG near the tachocline, again in good agreement with our result.
By contrast, substantially larger fields have been proposed in other contexts, including  $\sim1.5\times10^4$\,G required by mean-field flux-transport dynamo models \citep{Rempel2006}, $(4$--$10)\times10^4$\,G fields obtained in thin-flux-tube emergence simulations \citep{Fan2021}, and $(4$--$7)\times10^5$\,G estimates inferred from helioseismic analyses \citep{Chou2002, Baldner2008}. 
Reconciling these substantial discrepancies in the toroidal field strength estimates is likely an important topic for future investigation.

We further note that, in our interpretation linking the measured frequencies $\nu$ to slow magneto-Rossby modes, the fast modes are entirely absent from the observations.
These modes may have intrinsically weaker amplitudes, making them less effective at modulating magnetic flux emergence and therefore unlikely to leave a detectable imprint on AR emergence patterns.
Alternatively, their predicted frequencies, which are below $-100$\,nHz in our calculations, correspond to very long periods, so that their signatures may be smeared out over long-term temporal evolution and become difficult to identify observationally.

We also note that the discussion in this section is limited to establishing a connection between surface magnetic patterns and tachocline dynamics. 
We have not explored alternative scenarios, such as processes in the bulk of the convection zone or the near-surface shear layer, to explain the observed magnetic patterns. 
Therefore, we cannot rule out the possibility that our observations may also be explained by other mechanisms.

\subsection{Hemispheric Asymmetry}
\label{sec53}
In this study, we analyze the northern and southern hemispheres separately because of the clear north-south asymmetry in magnetic activity, evident in Figures~\ref{fig:long_sequence_N} and \ref{fig:long_sequence_S}.
Figures~\ref{fig:connct_lines_N} and \ref{fig:connct_lines_S} indicate that the dominant flux-integration bands in the two hemispheres appear to have substantially different relative rates, and Figure~\ref{fig:pow_NS} further suggests that the power spectra derived from the two hemispheres differ significantly. 
This raises the question of what causes these apparent hemispheric differences.

A closer examination of Figures~\ref{fig:pow_NS} and \ref{fig:ell_power}, however, indicates that the two hemispheres are not as different as they appear. 
Figure~\ref{fig:pow_NS} shows that the prograde $m$=1 modes in the two hemispheres have remarkably similar frequencies, but their power distributions differ. 
The main difference in magnetic-flux power arises in the retrograde modes: the northern hemisphere exhibits two prominent modes at $m$=1 and 2, whereas the southern hemisphere shows only a weaker mode at $m$=2.
Despite these differences, both the prograde and retrograde modes can be reconciled in Figure~\ref{fig:ell_power} under very similar physical conditions, provided that the angular velocity $\Omega$ in the tachocline of the southern hemisphere is about 1.4\,nHz slower than that of the northern hemisphere.

In other words, the physical conditions in the tachocline appear to determine the magneto-Rossby modes and their corresponding frequencies, whereas what determines the power amplitude of these modes remains unclear. 
It is this power distribution, along with the phases of each mode, as a function of $\ell$, $m$, and $\nu$, that ultimately governs the observed patterns of magnetic-flux emergence on the solar surface.
However, whether the physical conditions, such as rotational rates and toroidal field strengths at the tachocline, and the hemispherical asymmetries for different solar cycles are similar to Cycle 24 is an interesting question to investigate.

\section{Conclusions}
\label{sec6}

Our analysis of long-term spatiotemporal magnetic-flux maps offers a plausible explanation for the observed emergence and clustering patterns of major ARs and highlights their possible connection with the deeper solar interior. 
We have identified three spatiotemporal bands in each hemisphere, two of which rotate faster than the Carrington rate and one slower, that maximize the integration of total magnetic fluxes throughout the solar cycle.
The majority of major ARs emerge and cluster within or near these bands, particularly within or close to the intersections of two or three bands. 
These dominant flux bands have a one-to-one correspondence with $m$=1 modes (and in one case, $m$=2), which appear to correspond to slow magneto-Rossby waves that arise from the tachocline with a toroidal field strength around 4.0\,kG. 
The mode frequencies derived from the magnetic-flux spatiotemporal maps place a robust constraint on the physical conditions in the tachocline.
Although magneto-Rossby waves are unlikely to be responsible for generating the solar magnetic field, our results strongly indicate that they play a significant role in modulating when and where major ARs appear. 
While the Rieger-type periodicities likely arise from interactions between one dominant wave mode and other weaker modes, the quasi-periods of 0.6--4\,yr are likely produced by the intersections of two or three major wave modes.


\begin{acknowledgments}
\sdo\ is a NASA mission, and HMI is an instrument onboard \sdo\ and was developed by Stanford University under the NASA contract number NAS5-02139. 
This work is partly sponsored by NASA DRIVE Science Center COFFIES project under grant number 80NSSC22M0162.
The HMI magnetic field data are available from \url{http://jsoc.stanford.edu/} and the far-side helioseismic imaging data are available from \url{http://jsoc.stanford.edu/data/timed/}.
We thank Dr.~M.~Dikpati for reading our manuscript and providing constructive suggestions that have helped to improve the quality of this paper. 
\end{acknowledgments}

\bibliography{ms}{}
\bibliographystyle{aasjournal}

\end{document}